\documentclass[submission]{eptcs}
\providecommand{\event}{DACOMSIN 2021} 
\usepackage{graphicx}

\usepackage[sorting=none]{biblatex}

\bibliography{mambo_reference}

\title{Molecular and Materials Basic Ontology: development and first steps}
\author{Fabio Le Piane
\institute{Istituto per lo Studio dei Materiali Nanostrutturati (ISMN)}
\institute{Consiglio Nazionale delle Ricerche (CNR)\\
Bologna, Italy}

\email{fabio.lepiane@ismn.cnr.it}
\and
Matteo Baldoni
\institute{Istituto per lo Studio dei Materiali Nanostrutturati (ISMN)}
\institute{Consiglio Nazionale delle Ricerche (CNR)\\
Bologna, Italy}
\email{matteo.baldoni@ismn.cnr.it}
\and
Mauro Gaspari
\institute{Alma Mater Studiorum}
\institute{Universit\`{a} di Bologna\\ 
Bologna, Italy}
\email{mauro.gaspari@unibo.it}
\and
Francesco Mercuri*
\institute{Istituto per lo Studio dei Materiali Nanostrutturati (ISMN)}
\institute{Consiglio Nazionale delle Ricerche (CNR)\\
Bologna, Italy}
\email{francesco.mercuri@cnr.it}
}
\def\titlerunning{MAMBO: Materials And Molecules Basic Ontology}
\def\authorrunning{F. Le Piane, M. Baldoni, M. Gaspari, F. Mercuri}
\begin{document}
\maketitle
\begin{abstract}
Advanced materials and their applications have become a key field of research, and it looks like this trend is not going to change soon. For that reason, the need for systematic and efficient methods for organizing knowledge in the field and conduct computational or experimental investigations is stronger than ever.
In this work, we present a basic implementation of MAMBO - an ontology for molecular materials and their applications in real-life scenarios.
The development of MAMBO has been guided by the needs of the research community involved in the development of novel materials with functional properties, with particular attention to the nanoscale.
MAMBO aims at extending the current work in the field, while retaining a modular nature in order to allow straightforward extension of concepts and relations to neighboring domains. 
Our work is expected to enable the systematic integration of computational and experimental data in specific domains of interest 
(nanomaterials, molecular materials, organic an polymeric materials,
supramolecular and bio-organic systems, etc.). 
Moreover, MAMBO is developed with a strong focus on the applications of data-driven frameworks for the design of novel materials with tailored characteristics.
\\
\\
\textbf{Keywords:} Ontology; Materials Science; Nanomaterials; Molecular Materials; Knowledge Representation; Machine Learning
\end{abstract}

\section{Introduction}
The progress of a wide range of fields in science and technology
has greatly benefited from the development of new tailored
functional materials, addressing specific needs.
For that reason, advancements in materials development and manufacturing are considered key sectors for innovation and socio-economical assets\cite{KeyPolicy}.
Moreover, the recent developments of data-driven technologies led to significant progress in most strategic fields\cite{2021DataApplications,QIN2012220},
one of which is research and innovation for materials\cite{Himanen2019Data-DrivenPerspectives,Li2019ADomain,Pollice2021}.
Another piece of the puzzle is the amazing progress made in multiscale modelling and data-science approaches\cite{Agrawal2016Perspective:Science,LePiane2020PredictingLearning}, and the specific advancements in high-performance and high-throughput computing (HPC/HTC) and artificial intelligence served as a solid base for the applications of derived techniques.\\
The actual state-of-the-art approach 
for the design and development of novel materials
is based on tight integration between computational and experimental methods. Computational techniques are able to tackle a multitude of scenarios, from electronic structure calculations to continuum (full-scale) simulations\cite{Rosso2017WhatVersion}, while also giving the possibility to employ multi-scale techniques to link knowledge about materials spanning across a range of spatial and temporal scales. Moving to the experimental workflows, researchers often employ a variety of methodologies in order to gather information about materials during the entire development process. Both approaches share a trait: they are able to produce a large quantity of unstructured information on the objects they analyze and study. \\
Because of that, the dimension of data related to materials science increased enormously, leading to a strong need to organize and structure such information. Initiatives related to FAIR (Findable, Accessible, Interoperable, Reusable) requirements will further push the development on functional molecular materials.\cite{Wilkinson2016Comment:Stewardship}.
This strong need for organization can be
fulfilled by ontologies. While being still in its early stages in the domain of materials,
ontologies are already showing their great potential in the field\cite{Ashino2010MaterialsKnowledge,Cheung2008TowardsMaterials}.
In consequence of this emerging interest, recent research and cooperation activities have addressed the development of ontologies targeted to materials.
The creation of prolific platforms for data sharing in materials science is bound to the cooperation of group of researchers motivated to realize semantic technologies (like ontologies) able to unify and aggregate all the efforts and research lines already existing\cite{Horsch2020OntologiesMarketplace}. This is a crucial step on the road to enabling the digitalisation of materials design and development.
Indeed, we are already witnessing a huge amount of work in this direction, with particular focus on the development of top- and middle-level ontologies for materials science. A particularly relevant case is the European Materials Modelling Ontology (EMMO)\cite{Horsch2020ReliableOntology}. Stemming from this seminal effort, many domain ontologies tailored for specific use cases were born\cite{Li2020AnDomain, Li2019ADomain, Degtyarenko2008ChEBI:Interest}. 
However, for materials where aggregation properties at the molecular level are relevant, we can still face deficiencies in the development and application of structured knowledge.\\
MAMBO - the Materials and Molecules Basic Ontology - aims at filling this gap. 
MAMBO is focused on a specific domain related to materials science, which include, for example, molecular materials, nanomaterials, supramolecular materials, molecular thin-films and other similar systems. Many strategic fields like organic electronics and optoelectronics (OLEDs, organic thin-film transistors), organic and hybrid photovoltaics (organic and perovskite solar cells), bioelectronics (neural and brain interfaces), molecular biomaterials, and several others strongly depend on this kind of materials.
MAMBO is developed with modularity in mind, aiming to be easily extensible in order to cover other related aspects (i.e. a new computational method) or even to proximate domains. Morever, its concepts and relations are intended to be usable while defining new ontologies tailored to specific tasks and in order to provide full interoperability between different applications.\\
MAMBO is intended to lead to efficient data storage and retrieval infrastructures, merging information obtained via computational or experimental method with seamless transition from one to another. It can also provide the basis for a easier integration between data-driven technologies and classical materials science workflows. For example, machine learning based techniques for the design and development of novel functional materials would strongly benefit from a unification of knowledge on molecular materials and their 
representations.

\section{Related Work}
There are already different efforts in the field of ontologies for materials science domain focusing on different aspects and details. The already mentioned EMMO constitutes a significant example of a general ontology for the whole domain of materials modelling\cite{Ghedini2017EMMOONTOLOGY}, from which many others spawned focusing on specific use cases or on operational applications.
One example is ChEBI (Chemical Entities of Biological Interest), an ontology focused on chemical systems\cite{Degtyarenko2008ChEBI:Interest}. Another recent example is the Materials Design Ontology (MDO), an ontology defining concepts and relations useful to organize knowledge in the realm of materials design, with particular focus on solid-state physics\cite{Li2020AnDomain}.
It must be noted that despite a strong focus on specific use cases, concepts from these ontologies can in principle be reused in wider or slightly different domains. Accordingly many of the concepts we introduced in MAMBO are borrowed from ChEBI and MDO.\\
Moreover, even ontologies developed in other related domains (like digitalisation and virtualisation) can be related to MAMBO. Ontologies like OSMO (ontology for simulation, modelling, and optimization), and ontologies developed within the European project VIMMP (Virtual Materials Marketplace Project)\cite{Horsch2020OntologiesMarketplace,2020VIMMPMarketplace} also proved to be useful resources for re-using concepts, structures and relations.\\
Lastly, MAMBO also aims at connecting with pre-existing materials databases, like OPTIMADE and NOMAD\cite{2021TheOPTIMADE,2021TheNOMAD,Draxl2018NOMAD:Science}.

\section{Application scenarios}
MAMBO is tailored to the typical frameworks for the development of molecular materials and akin systems. In particular, we identified the following two main scenarios: i) retrieving structured information on molecular materials and ii) supporting the development of new, complex workflows for modelling systems based on molecular materials.\\
These can be complex tasks, where data can contain information about the basic entities that constitute parts of a target system (i.e. molecules, polymers, etc.). A good example is that of multi-scale modelling and characterization data on OLEDs, such as those discussed in \cite{Andrienko2020,Baldoni2018SpatialDynamics}. Another example use case for MAMBO could be the modelling of complex computational workflows for specific problems related to materials science. Moreover, MAMBO can help to organize the process of using data obtained by simulations in order to implement data-driven techniques in order to realize predictive models for tasks like property prediction, designing new materials and so on. This will also benefit from the semantic interoperability provided by MAMBO, which will give researchers the ability to integrate data between simulations and empirical experiments, further empowering data-based algorithms application.

\section{Development process, principles and methods}
The whole development process started with meetings with domain experts, aiming to define possible applications. These meetings allowed us to define:
\begin{itemize}
    \item A set of questions that MAMBO should answer (competency questions).
    \item A set of tasks that MAMBO should help to organize.
    \item A set of use cases.
\end{itemize}
Due to the peculiar nature of the typical development approaches pursued in the considered application area, we modelled the main concepts of the ontology associating them to specific problem solving methods (PSMs).
\cite{Fensel2003TheUPML}. This choice has been made because of the intrinsic peculiarity of the development processes considered.
Indeed, PSMs gives the possibility to define specific operations able to fulfill specific requirements and to reach the goals of a specific task. This is achieved by decomposing complex tasks into simpler subtasks, and then defining pre- and post-conditions for each of them. Thanks to this approach, we
were able to identify the indispensable terms needed to describe materials science, together with the connections that resides between different concepts stemming from such terms.\\
From the terms obtained in these first steps, an initial representation of the concepts and relations was drafted. In more detail, a “hybrid” approach (bottom-up and top-down) was used, to better represent the different nature of concepts involved in the development of the MAMBO ontology. A tentative set of relationships among terms was initially built and adjusted iteratively. Further details about the development process of MAMBO will be provided in a future work.
We then realized a first representation of the main concepts and their respective relations, drawing from the terms identified in the previous step, using a mix of bottom-up and top-down approaches, in order to better represent concepts from different scales and domains (i.e. computational and experimental workflows). This first scheme has then been adjusted iteratively.

\subsection{Integration with related ontologies}
Together with our specific design choices, we strove to keep MAMBO as integrated as possible with previous ontology related to the materials science domain. To this end, many terms introduced in MAMBO have been borrowed from such ontologies, which are:
\begin{itemize}
    \item EMMO - a standard representational ontology for applied science
    \item ChEBI - an ontology developed basing on a dictionary for "small molecular entities". The connection between MAMBO and ChEBI involves concepts related to individual molecules
    \item MDO - an ontology for materials design, with particular focus on crystalline structures. In MAMBO, we integrated the organization for crystals (usually inorganic) with our approach to molecular (organic) materials.
\end{itemize}
To enforce such integration, we started the conceptualization of MAMBO using components already defined in this existing work, while redefining those that needed to be adjusted to our peculiar domain. At the actual state, a better work on integrating MAMBO with these ontologies is still a work in progress.

\section{Realization of MAMBO}
We then proceed to define the core concepts and their mutual relationships, respecting the design principles described previously. We strove to give to MAMBO a modular structure in order to make it as easy to extend as possible in order to cover new domains and use cases.   

\subsection{Core concepts}
\begin{figure}[!ht]
	\centering
	\includegraphics[scale=0.3]{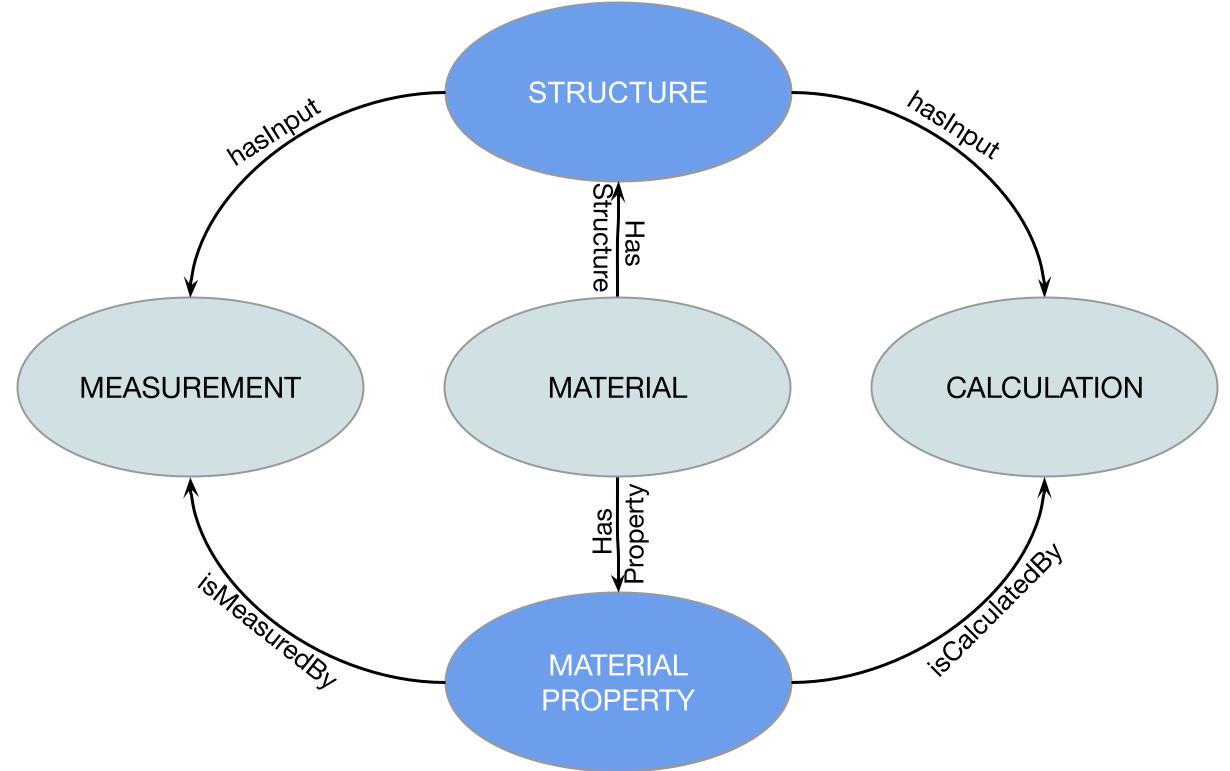}
	\caption{Core concepts of MAMBO: the ontology revolves around the concepts of {\fontfamily{tt}\selectfont Material}, {\fontfamily{tt}\selectfont Calculation} and {\fontfamily{tt}\selectfont Measurement}. An object ({\fontfamily{tt}\selectfont Material}) is represented by its structural features and properties.
	}
	\label{core}
\end{figure}

The very core of MAMBO includes the most fundamental terms stemmed from the aforementioned process, which will be represented by classes in the actual ontology. The general structure emerging is the following: 
\begin{itemize}
\item The central concept is that of {\tt Material}, which identifies the actual object of investigation
\item {\tt Material}s are defined mainly by their {\tt Structure}, which is the class containing the information about the structural characteristics of the material
\item {\tt Material}s have properties, which describes how they interact with the rest of the system/environment (and which are described in the {\tt Material Properties} class)
\item {\tt Material Properties} and {\tt Material Structures} can be the input and output of an experimental process (here {\tt Measurement}) or a computational one (here {\tt Calculation})
\end{itemize}
A scheme showing these concepts is shown in Fig. \ref{core}. 

\subsection{Drafting the "Structure" class}
\begin{figure}[!ht]
	\centering
	\includegraphics[scale=0.24]{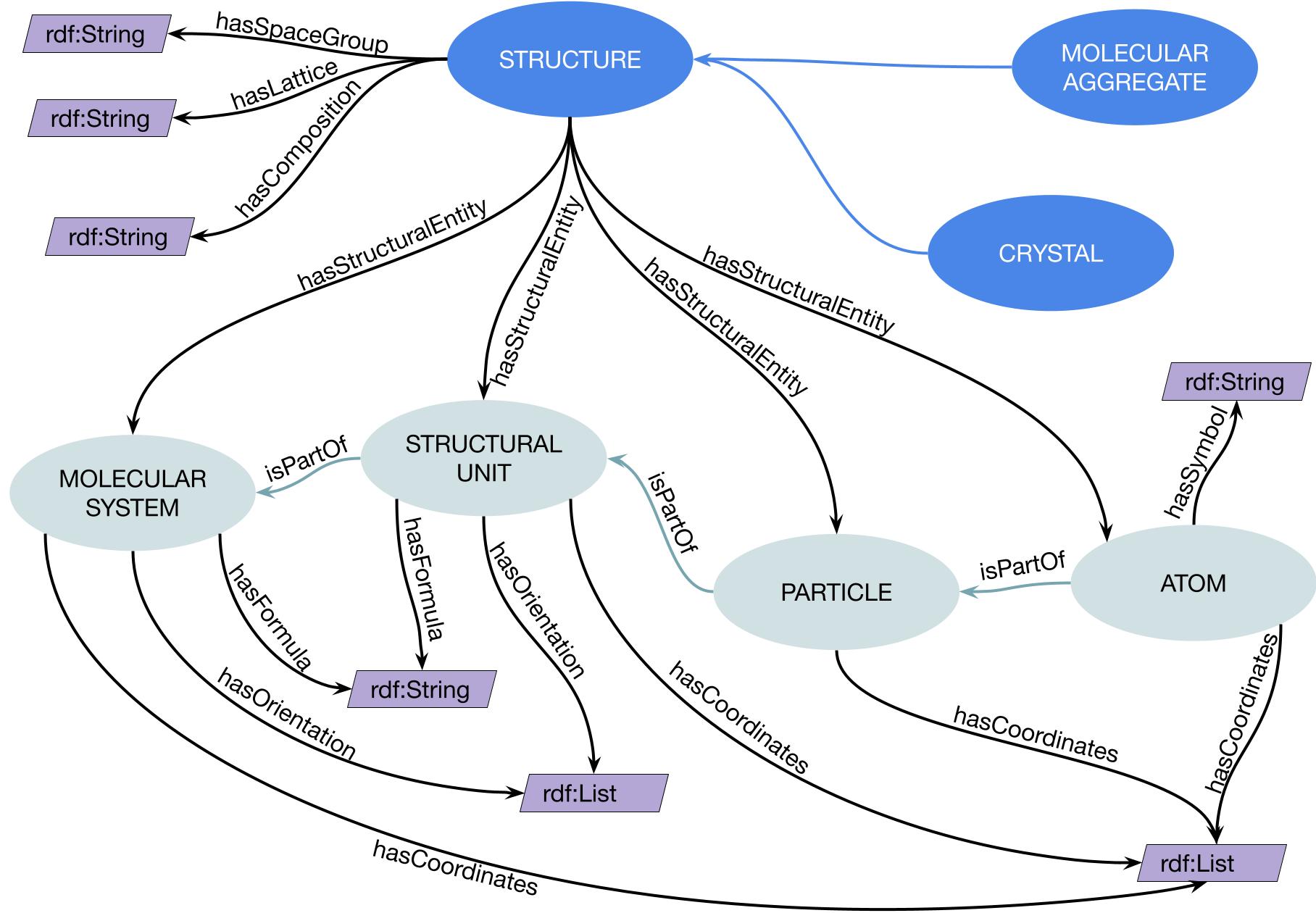}
	\caption{
	Draft scheme of the {\fontfamily{tt}\selectfont Structure} class. The main concepts and relationships used in the {\fontfamily{tt}\selectfont Structure} class are related to the analysis of actual workflows emerging from typical problem solving tasks involving molecular materials. Terms and relationships are connected to both computational and experimental techniques and methods.}
	\label{structure}
\end{figure}

We then proceeded to defining the {\tt Structure} class. The concepts and relationships identified at the time of this writing are shown in Fig. \ref{structure}. 
The figure shows all the concepts needed in order to organize the knowledge about the structure of molecular materials and systems based on them.\\
As already mentioned, the {\tt Structure} class role is to contain the information regarding the structural characteristics (in 3D space and time) of and object. The main choice we made in this realm is to describe a structure as composed by one or many ''structural entities'' having different features. Such structural entities could be atoms, particles, functional groups, molecules and so on. Morever, we defined focused subclasses of the {\tt Structure} class in order to represent more complex but fundamental systems like  {\tt Molecular Aggregate}s and  {\tt Crystal}s. Going further, we introduced more lower-level classes (in particular, we focused on reusing and extending concepts already present in related ontologies). For example, we introduced the concepts of position via the {\tt Coordinate} class (which can contain instances like center of mass, cartesian coordinates etc) and the concept of  {\tt Orientation} via the homonymous class (whose instances could be Euler angles, quaternion and rotation matrix). 
Finally, the {\tt Structure} class have properties related to the material in its integrity, like its periodicity.
For the sake of clarity, only a subset of all these concepts and relations are shown in Fig. \ref{structure}
.\\
As briefly mentioned, we also conducted brief instantiation tests, which ended up with promising results. Let us consider the case of a simulation of liposomes in water solution as an example. The main entity analyzed is the liposome structure, which is actually a lipid bilayer with a specific shape. It is straightforward to say that the liposome is going to be the instance of {\tt Molecular aggregate}, while the phospholipid which compose the liposome will be the instance of the {\tt Molecular System} class. Going forward, we can classify the molecule of the phospholipid as a {\tt Structural Unit}, having the related properties like charge. One of its phosphate group is and instance of the {\tt Particle} class and, finally, a phosphorus atom is easily assignable to the {\tt Atom} class. It should also be noted that the water surrounding the liposome (and the water actually contained within the liposome cavity) should be considered as a second instance of the {\tt Structure} class.

\subsection{The "Property", "Measurement" and "Calculation" classes}
We then shifted to the other core concepts of MAMBO, namely {\tt Property}, {\tt Measurement} and {\tt Calculation}, while also investigating their mutual releationships.\\
\begin{figure}[!ht]
	\centering
	\includegraphics[scale=0.5]{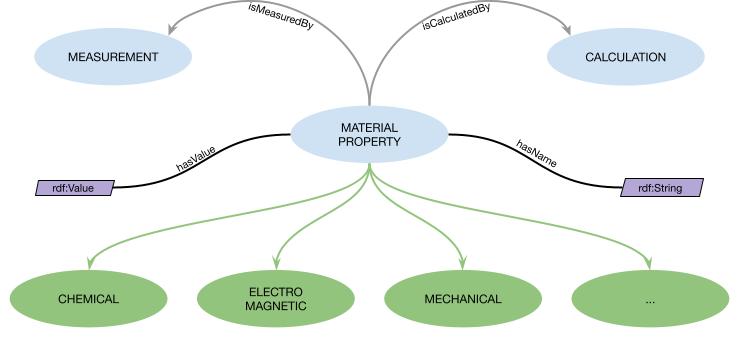}
	\caption{
	Scheme of the Property class. This class is linked to {\tt Measurement} and {\tt Calculation}. This connections and relations are developed in order to allow interoperability between experimental and computational workflows and data.{\tt Material Property} also has many subclasses related to different types of properties ({\tt Chemical}, {\tt Electromagnetic}, {\tt Mechanical} and so on).
	}  
	\label{property}
\end{figure}

\begin{figure}[!ht]
	\centering
	\includegraphics[scale=0.5]{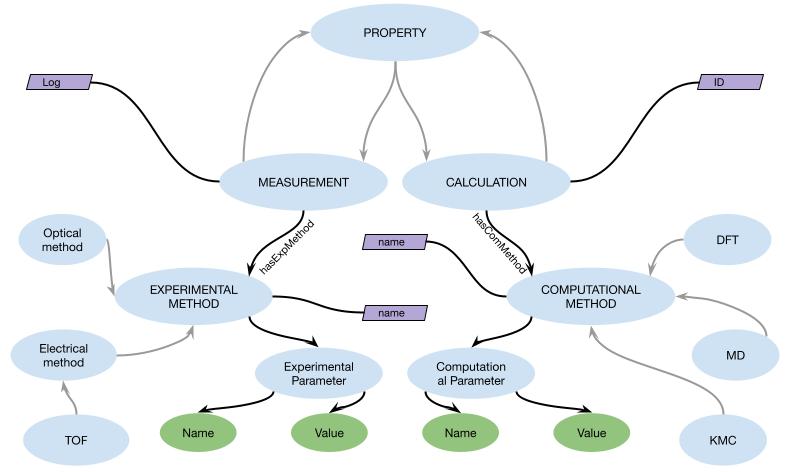}
	\caption{
	Scheme of {\tt Measurement} and {\tt Calculation}. They both have their respective ''method'' class, {\tt Experimental Method} and {\tt Computational Method}, respectively, which lead to the different experimental and computational methods, while gathering their parameters.
	}
	\label{meas_calc}
\end{figure}
These three classes are strongly interconnected (and also with the {\tt Structure} class): a {\tt Property} or a {\tt Structure} could be the results of a experimental measurement or of a computational workflow, respectively, represented by {\tt Measurement} and {\tt Calculation}. These last two classes are intended to be as similar as possible, meaning that the will have similar organisation and symmetrical relations with the other classes. This design is part of our strategy to make computational and experimental workflows as interoperable as possible, aiming to make data resulting from the twos easy to be leveraged together. At the same time, is important to be able to distinguish data and results coming from computational or experimental research, so both {\tt Measurement} and {\tt Calculation} have a specific class for their corresponding methods ({\tt Experimental Method} and {\tt Computational Method}) which are used to represent many different methodologies and their respective parameters. 
This organization is shown in Fig. \ref{property} and Fig \ref{meas_calc}.

\subsection{Formalization and implementation procedures}
We started developing the core and then moved to the other parts of the ontology. At the time being, the MAMBO core (Fig. \ref{core}) is implemented with the corresponding relations, and also {\tt Structure} and {\tt Property} general structure have been implemented but relations with their nested subclasses and other related classes are still a work in progress.
To this end, we used the OWL 2 language\cite{W3COWLWorkingGroup2012OWLOverview}\footnote{A draft version of the OWL implementation of MAMBO is
available on GitHub at: https://github.com/egolep/MAMBO}, 
while we are evaluating the possibility of re-implementing MAMBO with the OWLReady framework and library\cite{Lamy2017Owlready:Ontologies}. The RDF/XML syntax was used.\\
To implement MAMBO, we started by drawing the informal representation of a module, then trying to define the relations between the selected concepts, and finally identifying the main properties for each class and the corresponding subclasses. This also meant that we had to sketch the main hierarchies for these classes, and such hierarchies have been identified using the hybrid (i.e. both top-down and bottom-up) approached already discussed.
As expected, our reasoning turned out to be slightly imperfect. However, the main concepts the actual implementation are almost identical to that of the informal scheme. We pruned some hierarchies which proved to be redundant or misleading when used in practical cases. Such pruning led to fewer classes and shallower hierarchies, while enriching the relations aspect of MAMBO. 
At the time of this writing, we implemented all the main modules ({\tt Material}, {\tt Structure}, {\tt Material Property}, {\tt Measurement} and {\tt Calculation}), which are the same depicted in Fig. \ref{core}, Fig. \ref{structure}, Fig. \ref{property} and Fig. \ref{meas_calc}, respectively,  while some relations are still a work in progress. In particular, relations naming is susceptible to changes.
Other than working on a consistent naming standard, we are also proceeding with instantiation testing in order to see if the general structure still holds and to see which minor concepts we still need to add. For example, the {\tt Molecular Aggregate} and {\tt Crystal} classes have been added after a instantiation test that revealed their necessity. Major focus is now shifted to the {\tt Calculation} and {\tt Measurement} classes. 

\section{Future steps}
MAMBO is still under active and intense development, in particular we need to keep working on instantiation and modellation of real-world workflows in order to see if the implemented architecture holds.\\
While the core and the main concepts proved to be effective, the deeper hierarchies will probably require more care and are thus susceptible to changes. Moreover, a certain amount of work will also be needed in order to give consistent and proper naming to the relations used in order for MAMBO to be more easily understandable for domain experts. This process will also give us the opportunity to assess the effectiveness of different implementation strategies.\\
Then, our attention will shift to the extending MAMBO in order to cover specialized domain. In particular, we aim to use MAMBO to organize on a formal standpoint the computational and experimental knowledge gained through research on molecular materials in a as-unified-as-possible fashion. Because of that, MAMBO needs to address a broad range of concepts and their respective relations in subjects like multiscale computational modelling and experimental characterization for many specific class of materials. For this specific topic, it is fundamental to be able to easily and efficiently reuse the terminology coming from other ontologies without loosing the possibility to progressively add new ones for the most different use cases.
Finally, we would like to use MAMBO in order to design a database for molecular materials, giving researchers the power a semantic approach to realize complex and deep queries based on a flexible yet solid organization of knowledge of the field. 

\section{Conclusions}
In this paper we introduced MAMBO, a new ontology for molecular materials research and design both in the realm of computational and experimental workflows, striving to make the two fully interoperable.\\
The project yarn for being able to model a wide spectrum of concepts and relationships used in the filed of molecular materials, including methods  and approaches coming from disciplines like multiscale modelling. Giving a common interface for data coming from empirical and computational workflows will enable a full integration of such data, which would prove to be a great added value both for the creation of a database containing pre-existing data and for the application of data-driven techniques, like machine learning, which will give researchers the possibility to gather new information (and then, new data) at a more rapid pace. Moreover, the development approach used during the development of MAMBO is meant to allow the extension of the semantic asset towards related fields in the domain of molecular materials, and the concepts and relationships defined within MAMBO can also be easily reused while developing other top-level ontologies. \\
Initial assessment and instantiation tests demonstrate how the structure of MAMBO holds and allows for great expressivity and representability in the specific field of molecular materials and nanostructures.  The formal implementation is still a work in progress, in particular because we are trying to extend the scope of classes while testing performance in the intended use cases and applications.

\printbibliography
\end{document}